\documentstyle[12pt]{article}
\def\be{\begin{equation}}
\def\ee{\end{equation}}
\def\ba{\begin{array}{c}}
\def\ea{\end{array}}

\def\ben{$$}
\def\een{$$}
\begin{document}

\titlepage

 \begin{center}{\Large \bf
Quantum toboggans
 }\end{center}

\vspace{5mm}

 \begin{center}
Miloslav Znojil

\vspace{3mm}

 \'{U}stav jadern\'e fyziky AV \v{C}R,
250 68 \v{R}e\v{z}, Czech Republic

\vspace{5mm}

 {e-mail: znojil@ujf.cas.cz }

\end{center}

\vspace{5mm}

\section*{Abstract}

Among all the ${\cal PT}-$symmetric potentials defined on complex
coordinate contours $C(s)$, the name ``quantum toboggan" is
reserved for those whose $C(s)$ winds around a singularity and
lives on at least two {\em different} Riemann sheets. An enriched
menu of prospective phenomenological models is then obtainable via
the mere changes of variables. We pay thorough attention to the
harmonic oscillator example with a fractional screening and
emphasize the role of the existence and invariance of its
quasi-exact states for different tobogganic $C(s)$.

 \vspace{9mm}

\noindent
 PACS 03.65.Ge


 \begin{center}
\end{center}

\newpage

\section{Introduction}

In spite of an impressive mathematical universality of Quantum
Mechanics, many of its physical implementations refrain merely to
a quantization of a classical point particle. In particular (i.e.,
say, in one dimension and for a single particle), the idea of a
measurability of the coordinate $x$ leads immediately to the most
popular differential Schr\"{o}dinger equation
 \be
 -\frac{\hbar^2}{2m} \frac{d^2}{dx^2}\Psi({x})
 + V({x})\Psi({x})=E\,\Psi({x})\,
 \label{SE}
 \ee
defined over the real axis of coordinates $x \in I\!\!R$. In
addition, whenever we need that the energy remains observable, the
potential $V({x})$ is being assumed {\em real}.

Bender and Boettcher \cite{BB,BBL} initiated a small revolution by
conjecturing that the similar differential Schr\"{o}dinger
equations might remain phenomenologically useful even if the
coordinate-like variable ${x}$ itself {\em ceases } to be
observable. In the other words, they endorsed a new model-building
philosophy based on a tentative transition from the real line of
$x \in I\!\!R$ to the suitable {\em complex} contours,
 \be
 {x} \in C(s) \subset\,l\!\!\!C\,,\ \ \ \ \ \ \ \ \ \ \ \ \ s \in
 (-\infty,\infty)\,.
 \label{contour}
 \ee
They claimed that such a complexification of $x$ need not
necessarily be in conflict with the basic principles of quantum
mechanics. Quantitatively, they illustrated this possibility by
the (mainly, numerical and WKB) study of spectra of a family of
the ${\cal PT-}$invariant potentials of the generic power-law
multinomial form
 \be
 V(x) = \sum_\beta\,g_\beta\,(ix)^\beta\,\ \ \ \ \ \ \ \ \ \ \
 g_\beta \in I\!\!R
 \label{onestar}
 \ee
where the linear involution operator ${\cal P}$ represents parity
while the antilinear complex-conjugation involution ${\cal T}$
mimics time reversal \cite{BB}.

The energy spectrum appears to remain real for an unexpectedly
broad class of the ${\cal PT-}$symmetric potentials
(\ref{onestar}). In fact, the ``weakening" of the usual
Hermiticity of the Hamiltonian to the mere ${\cal PT-}$symmetry
may prove useful in formal sense. Thus, in ref. \cite{BBL} the
pair of complexifications (\ref{contour}) and (\ref{onestar}) has
been shown to {\em broaden} the class of the partially solvable
(usually called quasi-exact, QE) models~\cite{Ushveridze} which,
in the new setting, incorporates also the quartic-oscillator
family
 \ben
 V_{BB}(x) = a\,(ix) + b\,(ix)^2+c\,(ix)^3+(ix)^4\,.
 \een
In ref. \cite{quartic} we noticed that the latter QE construction
may find an efficient extension to the singular interactions
 \ben
 V_{Z}(x) = a\,(ix) + b\,(ix)^2+c\,(ix)^3+(ix)^4+
  e\,(ix)^{-1}+f\,(ix)^{-2}\,.
 \een
Now we intend to proceed one step further along this line.

In section \ref{I} we start by a brief review of the current
${\cal PT}-$symmetric way of dealing with the singular forces. For
illustrations of the most relevant technical details we pick up
the decadic-oscillator model of ref. \cite{decadic}.

In the next section \ref{II} we come with the idea of equivalence
between different singular potentials. This map is mediated by the
suitable, ${\cal PT}-$symmetry-preserving changes of the variables
which leave the {\em form} of the Schr\"{o}dinger equations
unchanged. This idea leads directly to our present constructive
approach to the toboggans in paragraph \ref{II.3}.

For the sake of clarity, we just give full details for a very
specific class  of potentials (``screened harmonic oscillators").
Only in discussion in section \ref{III} we then return back to a
broader context. We mention both some peculiarities of simpler
models (not so well suited for illustrating the subtleties of
tobogganic structures) and some complications related to the
higher-degree polynomial potentials (requiring a more or less
purely numerical treatment). We also briefly mention some recent
progress concerning the possible physical interpretation and
applicability of the general ${\cal PT}-$symmetric models.

Section \ref{Sigma} finally summarizes our present results.

\section{Boundary conditions, ${\cal PT}-$symmetric way \label{I}}

\subsection{Illustration: Asymptotics of decadic oscillators}

One of the most user-friendly guides towards complex contours
(\ref{contour}) is provided by the oscillators (\ref{onestar})
with $\max \beta = 10$,
 \be
 V(x) = x^{10} + {\rm asymptotically \ smaller\ terms}\,.
 \label{asydec}
 \ee
The standard asymptotic boundary conditions on the real axis may
be employed in the first step,
 \be
 \psi(x) = e^{-x^6/6+ {\rm asymptotically \ smaller\ terms}}\,.
 \label{minus6}
 \ee
We may re-parametrize the large values of $x = \varrho \exp
i\varphi \in C(s)$ by real $\varrho \gg 1$ and angle $\varphi \in
(0,2\pi)$ in order to see that
 \ben
 \psi(x) = \exp
 \left [-\frac{1}{6}\varrho^6 \cos 6\varphi
 + {\rm asymptotically \ less\ relevant\ terms}
 \right ]\,, \ \ \ \ \ \ \ \varrho \gg 1\,.
 \een
This implies that in the limit $|x| \to \infty$ our ``physical
prototype"  wavefunction (\ref{minus6}) vanishes whenever the
asymptotic angle $\varphi$ falls inside one of the six intervals
 \ben
 \Omega_{(first\ right)} =
 \left (
 -\frac{\pi}{2}
 +\frac{\pi}{12},
 -\frac{\pi}{2}
 +\frac{3\pi}{12}
 \right ), \ \ \ \ \
 \Omega_{(third\ right)} =
 \left (
 -\frac{\pi}{2}
 +\frac{5\pi}{12},
 -\frac{\pi}{2}
 +\frac{7\pi}{12}
 \right ), \ \ \  \ldots\,
 \een
which represent the wedges visible on the first Riemann sheet in
Figure~1 (where the upwards-running cut is assumed though not
indicated). In the ${\cal PT}-$symmetric context this observation
enables us, first of all, to choose the simplest curve $C(s) \neq
I\!\!R$ in the ``minimal" manner proposed by Buslaev and Grecchi
\cite{BG},
 \be
 C_{(BG)}(s) = s - i\,\varepsilon, \ \ \ \ \ \ \varepsilon > 0, \
 \ \ \ \ s \in I\!\!R\,.
 \label{straight}
 \ee
For small $\varepsilon$ it does not deviate too much from the real
axis $I\!\!R$.

Currently we are going to assume the analyticity of our potential
(\ref{asydec}) in the whole complex plane of $x$ admitting,
possibly, just a singularity at $x=0$ from which a cut is to be
oriented upwards. Then, an almost arbitrary smooth deformation
$C_{(third-third)}(s)$ of $C_{(BG)}(s)$ is admissible reflecting
just the uniqueness of the analytic continuation of the
wavefunction. The ends of all these deformations
$C_{(third-third)}(s)$ cannot leave the original pair of the
subscript-indicated asymptotic wedges of course.

A nontrivial change of the energy spectrum may be expected  when a
new curve $C(s) \neq C_{(third-third)}(s)$ is picked up as
connecting another pair of the wedges. Once we do so in the
left-right-symmetric (i.e., ${\cal PT}-$symmetric) manner, two new
possibilities emerge in Figure~1, viz., the downwards- or
upwards-bent integration contours $C_{(first-first)}(s)$ and
$C_{(fifth-fifth)}(s)$, respectively. The spectra of the
respective energies $E_{(first-first)}$, $E_{(third-third)}$ and
$E_{(fifth-fifth)}$ will be different in
general~\cite{Turbiner,RG}.

A less common possibility is encountered when our ansatz
(\ref{minus6}) is replaced by the alternative prescription
 \be
 \psi(x) = e^{+x^6/6+ {\rm asymptotically \ smaller\ terms}}\,.
 \label{plus6}
 \ee
The corresponding change of the menu of the eligible
decadic-oscillator boundary-condition wedges is displayed, in
Figure~2, on the same Riemann sheet as used in  Figure~1.  The two
topologically nonequivalent ${\cal PT}-$symmetric contours $C(s)$
will connect the second or the fourth left and right wedges.

\subsection{Quasi-exact states in the decadic model}

In extensive literature on QE bound states \cite{preUshveridze},
the majority of their relevant properties has already been
explored. In contrast,  ${\cal PT}-$symmetric quantum mechanics
still keeps some of its secrets so that the new QE constructions
are quite common there \cite{BBqes,Monou}. Once we restrict our
considerations to the polynomial version of our previous example
with ${x} \in C_{(third-third)}(s)$,
 \be
- \frac{d^2}{dx^2}\,\varphi(x)
 +\frac{L(
 L+1)}{{x}^{2}}
 \,\varphi({x})
 +
 \label{AHOdec}
 \left [
 {x}^{10} +
 g_8\,{x}^{8}+
 g_6\,{x}^{6}+
 g_4\,{x}^{4}
  +
 g_2\,{x}^{2}
  \right ]
 \,\varphi({x})=
 E\,\varphi({x})
 \,,
\label{SEd}
 \ee
we discover the existence of the two nonequivalent threshold
components of $\varphi({x})$ near the origin,
  \be
  \varphi({x}) = c_1
  \varphi_1({x})+c_2
  \varphi_2({x}), \ \ \ \ \ \ \ \ \
 \varphi_1({x}) \sim {x}^{-L}, \ \ \ \ \ \varphi_2({x}) \sim
 {x}^{L+1}.
 \label{obes}
 \ee
In order to circumvent this technical subtlety, we shall restrict
attention to the mere half-integer angular momenta $L$. This will
enable us to construct the QE bound-state solutions to eq.
(\ref{SEd}) by the method of ref.~\cite{decadic} incorporating
both the components (\ref{obes}) in a single QE ansatz at any
$L+1/2 = M =  1, 2, \ldots$,
  \be
 \varphi({x}) = \exp \left ( -\frac{{x}^6}{6} -\alpha\,\frac{{x}^4}{4} -
 \beta\, \frac{{x}^2}{2} \right )
 \,\sum_{n=0}^{N-1}\,h_n\,{x}^{2n-L}\,,
 \ \ \ \ \ \ \ \ \alpha=\frac{g_8}{2}, \ \ \ \beta =
 \frac{g_6-\alpha^2}{2}\,.
  \label{ansatz}
 \ee
The assumption of the existence of such a type of an
``exceptional" bound-state solution(s) represents the very core of
the concept of the QE solvability \cite{Ushveridze}. Tacitly, it
is assumed that the degree-of-polynomiality integer parameter $N
\geq 1$ is arbitrary and that all the non-QE normalizable
bound-state solutions, whenever needed, may be also constructed by
some numerical technique.

From eq. (\ref{ansatz}) one may extract all the QE-termination
conditions. The first one fixes $ {g_4}={g}_4(M,N)=2\alpha
\beta+2M -4N-2$ and enables us to transform our prototype
differential equation (\ref{AHOdec}) into the following finite set
of recurrences,
 \be
 A_n\,h_{n+1} + B_n\,h_{n} + C_n\,h_{n-1} + D_n\,h_{n-2} =0, \ \ \
 \ \ \ \ \ n = 0, 1, \ldots, N\
 \label{recurrences}
  \ee
with coefficients
 \be
\ba
 A_n=(2n+2)(2n+2-2M),\ \ \ \ \ \ B_n=E- \beta\,(4n+2-2M),\\
 C_n= \beta^2-{g_2} - \alpha\,(4n-2M) ,
 \ \ \ \ \ \ \
 D_n=4\,(N+1-n). \ea \label{sear}
 \ee
Let us merely briefly recollect some basic features of their
solution.

\subsection{Elementary QE families at the smallest $M$}

One of the most unpleasant features of the algebraic system of
$N+1$ equations (\ref{recurrences}) for $N$ coefficients $h_n$ is
that it is overcomplete and, hence, nonlinear. Fortunately, a
simplification occurs whenever $1 \leq M \leq N-1$ since one of
the important coefficients vanishes in such a case, $A_{M-1}=0$.
This means that an upper subsystem of eq. (\ref{recurrences})
acquires an $M-$dimensional matrix form,
  \be
\left(
 \begin{array}{ccccc}
B_0&A_0&&&\\
 C_1&B_1&A_1&&\\
   D_2&C_2&\ddots&\ddots&\\
    &\ddots&\ddots&B_{M-2}&A_{M-2}\\
 &&D_{M-1}&C_{M-1}&B_{M-1}
 \ea \right )
\left ( \ba h_0\\ h_1\\ h_2\\ \vdots \\ h_{M-1} \ea \right ) =0\,.
  \label{matriksmall}
 \ee
Once we satisfy its secular equation
  \be
   \det \left(
 \begin{array}{ccccc}
B_0&A_0&&&\\
 C_1&B_1&A_1&&\\
   D_2&C_2&\ddots&\ddots&\\
    &\ddots&\ddots&B_{M-2}&A_{M-2}\\
 &&D_{M-1}&C_{M-1}&B_{M-1}
 \ea \right )
=0
  \label{eq4.8}
 \ee
one of the first $M$ lines of eq. (\ref{recurrences}) may be
omitted as linearly dependent. This is precisely what we need.
After {\em any such} omission, all our remaining recurrences
become tractable as another linear and homogeneous matrix problem,
solvable by the standard numerical techniques at any finite
dimension $N$~\cite{Wilkinson}.

The practical solution of our two coupled matrix equations will be
significantly facilitated whenever the ``smaller" eq.
(\ref{matriksmall}) remains solvable in closed form. Of course,
the constraint (\ref{eq4.8}) remains trivial and gives the unique
QE-compatible energy $E=0$ at $M=1$. An exhaustive analysis of the
related $N-$dimensional QE conditions has been given in ref.
\cite{decadic}. It has been emphasized that the QE wavefunctions
form the so called Sturmian $N-$plets at any positive integer $N$.
The corresponding QE-compatible couplings $g_2$ remain real (and,
hence, compatible with the physical ${\cal PT}-$symmetry
requirement) in a fairly large domain of the two freely variable
couplings $g_6$ and $g_8$.

Whenever we decide to omit the {\em first} linearly dependent line
from eq. (\ref{recurrences}), the resulting reduced secular
equation contains $g_2$'s just on the main diagonal. Of course, at
any $M \geq 2$, the apparent linearity of such a problem is lost
due to the nontrivial $g_2-$dependence of the roots of the
``auxiliary" constraint (\ref{eq4.8}). Fortunately enough, once
written in the form
 \ben
 {g_2}=\frac{E^2}{4}, \ \ \ \ \ \ M=2\,,
 \een
the solution of eq. (\ref{eq4.8}) remains sufficiently compact and
$N-$independent at $M=2$.

We see that with the growth of $M$, the determination of our
Sturmian $N-$plets of the QE-compatible energies and couplings
becomes more complicated, though not at a really prohibitive rate.
Even in the next step with $M=3$, the formula for $g_2$ remains
fairly similar and linear though $N-$dependent,
 \ben
g_2 = \frac{8}{E}\left (2 N-2-\alpha\beta \right )
+\frac{E^2}{16}\,,\ \ \ \ \ \ \ \  M =3\,.
 \een

\section{Transformations changing $C(s)$ \label{II}}

The key technical ingredient with which we intend to present
quantum toboggans will be certain equivalence transformation
between different ${\cal PT}-$symmetric models. Let us first
assume that the Schr\"{o}dinger eq. (\ref{SE}) + (\ref{onestar})
has the following slightly more specific, perturbative,
anharmonic-oscillator form
 \be
 \left [
 -\frac{d^2}{dx^2} - (ix)^2 + \lambda\,W(ix)
  \right ]
 \,\psi
 (x)=
 E(\lambda)
 \,\psi
 (x)\,,
 \ \ \ \ \ \ W(ix)=\sum_\beta\,g_\beta(ix)^\beta\,.
 \label{AHO}
 \ee
Many phenomenological considerations were based on such a type of
models in the past~\cite{anharm}. People usually assume that
$\lambda$ is ``sufficiently small" so that the reliable
approximations of the energies pertaining to eq. (\ref{AHO}) may
be calculated using perturbation expansions in the powers of
$\lambda$. Our present approach will be different.

\subsection{Liouvillean changes of variables
\label{II.1} }

Let us try to change the variables in eq. (\ref{AHO}) in the
manifestly ${\cal PT}-$symmetric manner,
 \be
 ix = (iy)^\alpha\,, \ \ \ \ \ \ \
 \psi(x) = y^\varrho\,\varphi(y).
 \label{change}
 \ee
This means that at any real exponent $\alpha>0$ we have
 \ben
 i \,dx = i^\alpha \alpha y^{\alpha-1}\,dy
 , \ \ \ \ \ \ \ \
  \frac{(iy)^{1-\alpha}}{\alpha}\,\frac{d}{dy}= \frac{d}{dx}\,.
  \een
Our ``old" Schr\"{o}dinger equation (\ref{AHO}) acquires the
``new", mathematically equivalent form
 \ben
 y^{1-\alpha}
 \frac{d}{dy}
 y^{1-\alpha}
 \frac{d}{dy}\,y^\varrho\,\varphi(y)
 +i^{2\alpha} \alpha^2 \left [
 - (iy)^{2\alpha} +
 \lambda\,W[(iy)^{\alpha}] -E(\lambda)
  \right ]
 \,y^\varrho\,\varphi(y)=0\,.
 \een
Its first term is a sum
 \ben
y^{1-\alpha}
 \frac{d}{dy}
 y^{1-\alpha}
 \frac{d}{dy}\,y^{[(\alpha-1)/2]}\,\varphi(y)
 = y^{2+\varrho-2\alpha} \frac{d^2}{dy^2}\,\varphi(y)
 +\varrho(\varrho-\alpha)y^{\varrho-2\alpha}\,\varphi(y)\,, \ \ \
 \varrho = \frac{\alpha-1}{2}\,
 \een
where the first derivatives of $\varphi(y)$ dropped out after our
choice of the value of $\varrho$ (this idea dates back to
Liouville \cite{LG}). Thus, the new Schr\"{o}dinger equation
preserves the standard form,
 \be
- \frac{d^2}{dy^2}\,\varphi(y)
 +\frac{\alpha^2-1}{4y^{2}}
 \,\varphi(y)
 +(iy)^{2\alpha-2} \alpha^2 \left [
 - (iy)^{2\alpha} +
 \lambda\,W[(iy)^{\alpha}] -E(\lambda)
  \right ]
 \,\varphi(y)=0\,.
 \label{AHOnew}
 \ee
It naturally contains the $\alpha-$dependent singularity so that
it does not make any sense to insist on its absence in our initial
problem~(\ref{AHO}).

\subsection{The screening choice of $W(ix)$
\label{I.1b} \label{I.2}}

It is well known that the requirement of a {\em feasibility} of
the evaluation of the individual perturbative corrections leads to
the preference of integers $\beta = 3, 4, \ldots$ \cite{Fluegge}.
In parallel, the requirement of the {\em convergence} of the
infinite perturbation series may only be met when operators or
functions $W$ are relatively bounded \cite{Kato}. In application
to our particular ansatz (\ref{AHO}) this means that we must
consider only $\beta < 2$.

One of the ways out of such a trap lies in a transition to
semi-numerical strong-coupling expansions \cite{UshTur}, a
remarkable technical simplification of which is easily found to
occur at the {\em rational} exponents $\beta$~\cite{rnad}. This
motivates our present specific choice of
 \be
 W(ix)=
 a\,(ix)^{4/3}+
 b\,(ix)^{2/3}+
 c\,(ix)^{0}+
 d\,(ix)^{-2/3}+
 e\,(ix)^{-4/3}+
 f\,(ix)^{-2}\,
 \label{exponents}
 \ee
parameterized by the six real coupling constants. Note that the
exponents $\beta$ are {\em both} rational and smaller than two
here. In a {phenomenological} perspective, such a function seems
suited to screen the original oversimplified force $x^2$ in both
the short- and long-distance regimes.

Once we pick up the specific exponent $\alpha = 3$ in
(\ref{change}) and scaling $\lambda = 1/9$ in eq. (\ref{AHO}), our
Schr\"{o}dinger equation (\ref{AHOnew}) + (\ref{exponents})
precisely coincides with the decadic oscillator prototype problem
(\ref{SEd}). This is one of our present key observations. One is
only required to abbreviate $ L(L+1)\equiv 2-f$ (or, if you wish,
$L=L(f)= \sqrt{9/4-f}-1/2$) and to perform a series of the
following identifications: $g_8 \equiv a$, $g_6 \equiv -b$, $g_4
\equiv c-E(1/9)$, $g_2 \equiv -d$ while $E \equiv -e$. All these
formulae express just a freedom of the mathematical identification
of our two {\em phenomenologically different} decadic and screened
harmonic oscillators.

It is quite amazing that their one-to-one equivalence mapping
requires just an appropriate re-interpretation of the constants.
We are now, at last, prepared to discuss the particular
implications of this equivalence on the related,  topologically
nontrivial changes of the ``coordinate" contours $C(s)$.

\subsection{The emergence of tobogganic contours \label{II.3}}

The introduction of a complex deformation $ I\!\!R \to C(s)$ of
the domain of $x$ is a mere trivial analytic-continuation trick
for mathematicians. In the context of physics, nevertheless, it
requires an explanation \cite{Intro}. One may, for example,
re-classify the complex coordinate $x$ as a mere auxiliary
variable in a representation of the Hilbert space of states
\cite{BBJ}.  In some physical systems, one may also try to
construct a suitable, usually quite complicated operator of
certain quantity interpreted as an ``observable of
position"~\cite{Batal}.

In both of these two situations we are entirely free to adapt the
choice of $C(s)$ to our phenomenological needs emphasizing, for
example, an expected impact of the choice of the wedges upon the
spectrum of the observed energies. As we have emphasized in the
Introduction, the ``topologically trivial" contours (which may be
depicted in a single Riemann sheet) look, paradoxically,
intuitively less difficult to grasp and accept. Even then, we
might re-read Figures~1 and 2 as a demonstration that there exists
a wealth of the non-equivalent boundary conditions even in such a
simpler case. An even stronger support of such a point of view
might be found in the asymptotically exponential potentials of
refs. \cite{RG,CJT}. With their infinitely many (sic!) alternative
wedges (which are {\em all} visible on the first Riemann sheet)
they represent a certain quite difficult mathematical challenge,
say, when treated in the so called Stokes-geometry
language~\cite{Tanaka}.

The main purpose of our present paper is to analyze the contours
$C(s)$ which would become topologically nontrivial. Without any
real loss of generality, we may start from the the most elementary
choice of $\max \beta=2$ in our particular model
(\ref{exponents}). In such a case with the harmonic-oscillator
asymptotics $\psi(x) \sim  \exp (-x^2/2)$, there only exist the
two ``standard" harmonic-oscillator wedges which are visible on
the first Riemann sheet of Figure 3. In this picture with a branch
singularity at $x=0$, the complex plane is cut upwards.

Figure 3 also displays the first example of a quantum-toboggan
contour $C(s)$ which leaves the  first Riemann sheet. In fact,
many different quantum toboggans will share the part which is
visible in Figure 3. Once these contours $C(s)$ remain ${\cal
PT}-$symmetric, they will always connect the {\em same} wedges.
Thus, in our notation we shall have $C_{(third-third)}(s)$
connecting the third left with the third right wedge, etc.

In Figure 4 we return to the contour $C_{(second-second)}(s)$
which is derived from the alternative asymptotics $\psi(x) \sim
\exp (+x^2/2)$ and which, in essence, remains non-tobogganic. It
is still worth noting that its ``tobogganic" version could be
obtained by the mere analytic continuation since the picture is
able to display just the halves of the relevant wedges. The second
interesting observation is that the curve $C(s)$ cannot be
deformed to any vicinity of the real line - similar shape is also
exhibited by the Coulomb/Kepler contours~\cite{coul}.

The first nontrivial toboggan appears in Figure 5 where the
``left" (!) branch of the toboggan  $C_{(third-third)}(s)$ is made
visible by the clockwise rotation of the cut of Figure 3  by 90
degrees. The picture is made complete by Figure 6 where the
remaining ``right" half of the ``third-third"  toboggan is made
visible by the anticlockwise rotation of the cut of Figure 3  by
90 degrees.

In our final Figure 7, a ``left" branch of the fourth-fourth
toboggan is displayed. It corresponds to the ``anomalous"
asymptotics $\psi(x) \sim \exp (+x^2/2)$ and is made visible by
the clockwise rotation of the cut of Figure 3 by 180 degrees. The
visualization of the rest of this curve $C(s)$ is trivial as it
would just require an opposite, anticlockwise rotation of the cut.

\section{\label {III} Discussion}

In spite of their entirely natural character, the studies of all
the specific ``far-from-the-real-axis" contours $C(s)$ are by far
not frequent in the literature on ${\cal PT}$ symmetry even if we
restrict our attention just to the QE context
\cite{quartic,decadic,Monou}. In such a setting, our present
further move towards tobogganic $C(s)$ might look almost like a
heresy. At present we see at least two reasons for an appeal of
such a study. Firstly, we have already demonstrated that the
Liouvillean changes of variables might make the {\em explicit
constructions} of toboggans quite straightforward. Secondly, we
believe that a particularly appealing aspect of {\em all} the
modifications of boundary conditions is {\em revealed} by the QE
solvable models. They represent very specific systems where just
{\em some} of the levels remain unchanged due to the elementary,
exceptionally easily analytically continued form of their QE
wavefunctions.

\subsection{Back to the simpler models
\label{I.1a}}

For the above-mentioned reasons we tried to keep our overall
considerations in close contact with their specific
screened-harmonic-oscillator illustration. Now, it is time to
notice that the winding-independent coincidence of the energy
levels must occur also in the simpler solvable potentials.

\subsubsection{The unscreened harmonic oscillator}

It is well known \cite{ptho} that the harmonic-oscillator
Schr\"{o}dinger equation
 \ben
 \left (
 -\frac{d^2}{dz^2} + \frac{\alpha^2-1/4}{z^2} + z^2
 \right )
 \,h_n^{(\pm)}
 (z)=
 E_n^{(\pm)}\,
 \,h_n^{(\pm)}
 (z),\ \ \ \ \ \ \ z \in I\!\!R
 \een
is exactly solvable in terms of Laguerre polynomials $L_n^{(\pm
\alpha)}(z)$ at $\alpha = 1/2$,
 \ben
 E_n^{(\pm )}=4n+2\pm 2 \alpha, \ \ \ \ \
 \,h_n^{(\pm)}
 (z)= c_n^{(\pm)}\,\sqrt{z^{1\pm 2\alpha}}\,e^{-z^2/2}\,
 L_n^{(\pm\alpha)}(z^2),
 \ \ \ \ \ \ \ n = 0, 1, \ldots\,.
 \een
As long as these solutions are analytic in the whole complex plane
of $z$, the model offers in fact one of the simplest possible
illustrative examples of the existence of the real spectrum of
energies generated by a manifestly non-Hermitian, ${\cal
PT}-$symmetric Hamiltonian at $\alpha = 1/2$ \cite{BB}.

At $\alpha \neq 1/2$, the model gets regularized on the straight
line (\ref{straight}) at any $\varepsilon > 0$. One arrives at a
merely inessentially modified ordinary differential equation
 \be
 \left (
 -\frac{d^2}{dx^2} + x^2 - 2i\varepsilon\,x+
  \frac{\alpha^2-1/4}{x^2 - 2i\varepsilon\,x-\varepsilon^2}
 \right )
 \,h_n^{(\pm)}
 (x-i\varepsilon)=
 \left (
 E_n^{(\pm)}+\varepsilon^2\right )
 \,h_n^{(\pm)}
 (x-i\varepsilon)
 \label{HOshi}
 \ee
where $x \in I\!\!R$. In contrast to the original real-line
problem where the parity has been conserved, ${\cal
P}\,h_n^{(\pm)}(z) = h_n^{(\pm)}(-z)= \mp h_n^{(\pm)}(z)$, the new
eigenfunctions $h_n^{(\pm)}(x-i\varepsilon)$ only remain
eigenstates of the product of the two (mutually commuting)
operators ${\cal P}$ and ${\cal T}$. {\em Their} ${\cal
PT}-$eigenvalue might give a new meaning to the  superscripts of
$h_n^{(\pm)}(x-i\varepsilon)$ but one usually opts for complex
normalization constants $c_n^{(\pm)}$ chosen in such a way that
all ${\cal PT}-$eigenvalues of $h_n^{(\pm)}(x-i\varepsilon)$
become the same and equal, say, to one \cite{Ques}.

Of course, the analytic continuation will work at all the
accessible Riemann sheets so that {\em all} the tobogganic
versions of the harmonic oscillator will have the same spectrum.
In this sense, the example itself is degenerate and trivial.

\subsubsection{The Magyari's family of quasi-exact toboggans }

In QE models, only the levels with QE structure are easy to study.
Their formal resemblance to harmonic-oscillator states implies, in
particular, that the {\em different} physical situations may be
represented by the {\em same} formula since, as we have seen, the
{\em single} QE bound state may fit {\em several} boundary
conditions at once.

We have shown that once we fixed our ``prototype" decadic
polynomial problem (\ref{SEd}), we were able to map some of its
non-tobogganic versions upon the tobogganic variants of our
illustrative harmonic oscillator screened by our specific six-term
perturbation (\ref{exponents}). For pedagogical reason we choose
such a set of examples because, on one side, the Magyari's
\cite{Magyari} entirely general QE and ${\cal PT}-$symmetric
prototype problem
 \be
- \frac{d^2}{dx^2}\,\varphi(x)
 +\frac{L(
 L+1)}{{x}^{2}}
 \,\varphi({x})
 +
  \left [
 {x}^{4q+2} +
 g_{4q}\,{x}^{4q}+
 \ldots
  +
 g_2\,{x}^{2}
  \right ]
 \,\varphi({x})=
 E\,\varphi({x})
 \,
\label{SEget}
 \ee
is only exceptionally tractable non-numerically at $q>2$
\cite{Gerdt}. We were still able to recollect a few explicit
formulae for $q=2$ \cite{decadic}. On the other side, there
obviously exist also the simpler and more popular QE constructions
at $q=1$. We felt that they do not offer a sufficiently
transparent and persuasive introduction of a sufficiently large
variety of tobogganic paths $C(s)$. Of course, having now gained
the overall experience at $q=2$, we might easily return to the
multinomial $q=1$ models and study, say, the set of simple QE
examples
 \ben
 V_f(x)=
 {x}^{6} +
 f_{4}\,{x}^{4}+
  f_2\,{x}^{2}+
 f_{-2}\,{x}^{-2}, \ \ \ \ \
 V_g(x)=
 -({ix})^{2} +
 i\,g_{1}\,{x}+
  g_{-1}\,({ix})^{-1}+
  g_{-2}\,({ix})^{-2},
  \een
  \ben
  \ \ \ \ \
 V_h(x)=
 -({ix})^{2/3} +
 h_{-2/3}\,({ix})^{-2/3}+
  h_{-4/3}\,({ix})^{-4/3}+
  h_{-2}\,({ix})^{-2},
 \een
by choosing $\alpha=q+1=2$ for $V_f$ or $\alpha =  1$ for $V_g$ or
$\alpha =  2/3$ for $V_h$  in eq. (\ref{AHOnew}). Expecting no
really new observations, we leave this analysis as an exercise to
the reader.

\subsection{Physics with complex pseudo-coordinates $x \in C(s)$ }

Our study is based on the expected compatibility of a change of
the integration path $C(s)$ (and of the resulting loss of the
Hermiticity of the Hamiltonian $H$) with the observability (i.e.,
reality) of the energies. For the sake of completeness, let us now
briefly summarize that a tenable physical background of such a
construction requires that

\begin{itemize}

\item
for all the operators ${\cal O}$ in our Hilbert space ${\cal L}$
we replace the usual conjugation ${\cal O} \to {\cal O}^\dagger$
by an alternative involutive operation
 \be
 {\cal O} \to {\cal O}^\ddagger = \eta^{-1} {\cal O}^\dagger \eta
 \label{define}
 \ee
where the positive definite operator $\eta \neq I$ represents a
certain non-standard metric in our Hilbert space. In this
language, let the family of the observables be characterized by
the so called quasi-Hermiticity property,
 \be
 {\cal O}^\dagger = \eta {\cal O}\eta^{-1}\,
 \label{nehermiticity}
 \ee
(see an older general review \cite{Geyer} for more details);

\item
in the next step let us admit that eq. (\ref{nehermiticity})
(taken as an implicit definition of an unknown $\eta$ from a given
Hamiltonian ${\cal O} = H$) has more solutions. One of them is
assumed to coincide with the parity operator ${\cal P}$. In this
manner we quite naturally arrive at the concept of the ${\cal
PT}-$symmetric quantum Hamiltonians of ref. \cite{BB};

\item
at the same time we must be permitted to represent {\em all} the
observables as operators compatible with the original
quasi-Hermiticity condition (\ref{nehermiticity}). In this sense,
the complex coordinates are in general not observable.

\end{itemize}

 \noindent
In terms of physics, we have to proceed in opposite direction. The
standard probabilistic physical interpretation of the ${\cal
PT}-$symmetric theory remains only achieved  {\em after} our
construction of the modified scalar product mediated by the (not
necessarily unique \cite{Geyer}) metric operator
$\eta=\eta^\dagger > 0$. This operator must satisfy all the subtle
mathematical conditions of quasi-Hermiticity (cf.
\cite{Geyer,Ali}). For physical reasons, people often denote $\eta
\equiv {\cal CP}$ and speak about a charge-symmetry operator
${\cal C}$ and about the fundamental ${\cal CPT}-$symmetry in
field theory \cite{BBJ}.

In our present tobogganic context, we may conclude that one is
allowed to introduce complex ``coordinates" $x \notin
(-\infty,\infty)$ {\em provided only} that they are not treated as
``measurable" \cite{Introdu}. The spectrum of the related
non-Hermitian Hamiltonians $H \neq H^\dagger$ (i.e., by
assumption, the {\em observable} values of the bound-state energy
levels) {\em must} remain {real} of course.

One of the important open questions is {\em how} one could
guarantee this reality but, of course, this problem is shared by
both the tobogganic and non-tobogganic models.

\section{Summary \label{Sigma}}

Among innovations accepted in the context of the ${\cal
PT}-$symmetric version of quantum mechanics, one of the most
unusual ones concerns the admissibility of the various complex
contours of the (by assumption, unmeasurable) ``coordinates" $x
\in C(s)$. This possibility is comparatively rarely emphasized in
the related literature, presumably, due to the expected purely
technical complications. Only very recently, A. Mostafazadeh
\cite{sesiti} reminded us that such an expectation would be
over-pessimistic since all the ${\cal PT}-$symmetric (i.e.,
left-right symmetric) contours $C(s)$ in the complex plane of $x$
may be perceived simply as their properly matched left and right
sub-contours.

Whenever the potential in our Schr\"{o}dinger equation contains a
centrifugal-like singularity as contemplated, probably for the
first time, by Buslaev and Grecchi in ref. \cite{BG}, the
situation remains very similar - in the complex plane of $x$, it
is only necessary to introduce a (say, upwards-running) branch cut
from zero to infinity (cf., e.g., Figure 3 above). In our paper we
tried to persuade the readers that in such a situation, the {\em
existence} of the centrifugal-like singularity {\em simplifies}
the picture since it {\em allows} us to perform a fairly general
change of variables (cf. eq. (\ref{change}) in our present
context). This transformation mediates a {\em one-to-one
correspondence} between the models defined over {\em different}
contours of coordinates $C(s)$. In our considerations we
emphasized that many ${\cal PT}-$symmetric contours may be
visualized as maps of some other contours lying sufficiently close
to the real axis.

Our main use of the latter correspondence between the ${\cal
PT}-$symmetric integration contours pertaining to the different
potentials lied in the use of the mapping of the ``elementary"
contours $C_{(e)}(s)$ (i.e., the ones confined to a single Riemann
sheet) onto the ``tobogganic" contours $C_{(t)}(s)$ (i.e., the
ones which encircle the branch points and become defined over {\em
several} Riemann sheets in general).

In spite of a certain counter-intuitive character of the
tobogganic case, one should keep in mind that {\em both}
$C_{(e)}(s)$ and $C_{(et)}(s)$ are {\em equally} counter-intuitive
since, as we already mentioned (cf. also \cite{Introd}), our
``coordinates" $x \in C$ do {\em not} represent a physically
measurable quantity. Hence, there is no reason for omitting the
topologically nontrivial ``toboggan-shaped" curves $C_{(t)}(s)$
from ${\cal PT}-$symmetric quantum mechanics.

In our paper we showed that the use of the simplest versions of
the toboggans  $C_{(t)}(s)$ need not necessarily lead to any
perceivable technical difficulties. As long as we paid attention
just to the simplest possible class of $V(x)$ (possessing just the
single strong singularity at $x=0$), our ``recipe" of an
interpretation of the bound states degenerated simply to the
inverse Liouvillean change of variables.

In the conclusion, let us express our belief that even the ``next
move" to the study of some multiply tobogganic ${\cal
PT}-$symetric potentials (containing more than one strong
singularity) need not still lead to unsurmountable difficulties:
As an encouragement, one might cite, say, the existence of the
nice and solvable potentials of this type described in
ref.~\cite{Anjana}.

\section*{Acknowledgement}

Work supported by the Institutional Research Plan AV0Z10480505 and
by the GA AS CR grant Nr. A1048302.

\section*{Figure captions}

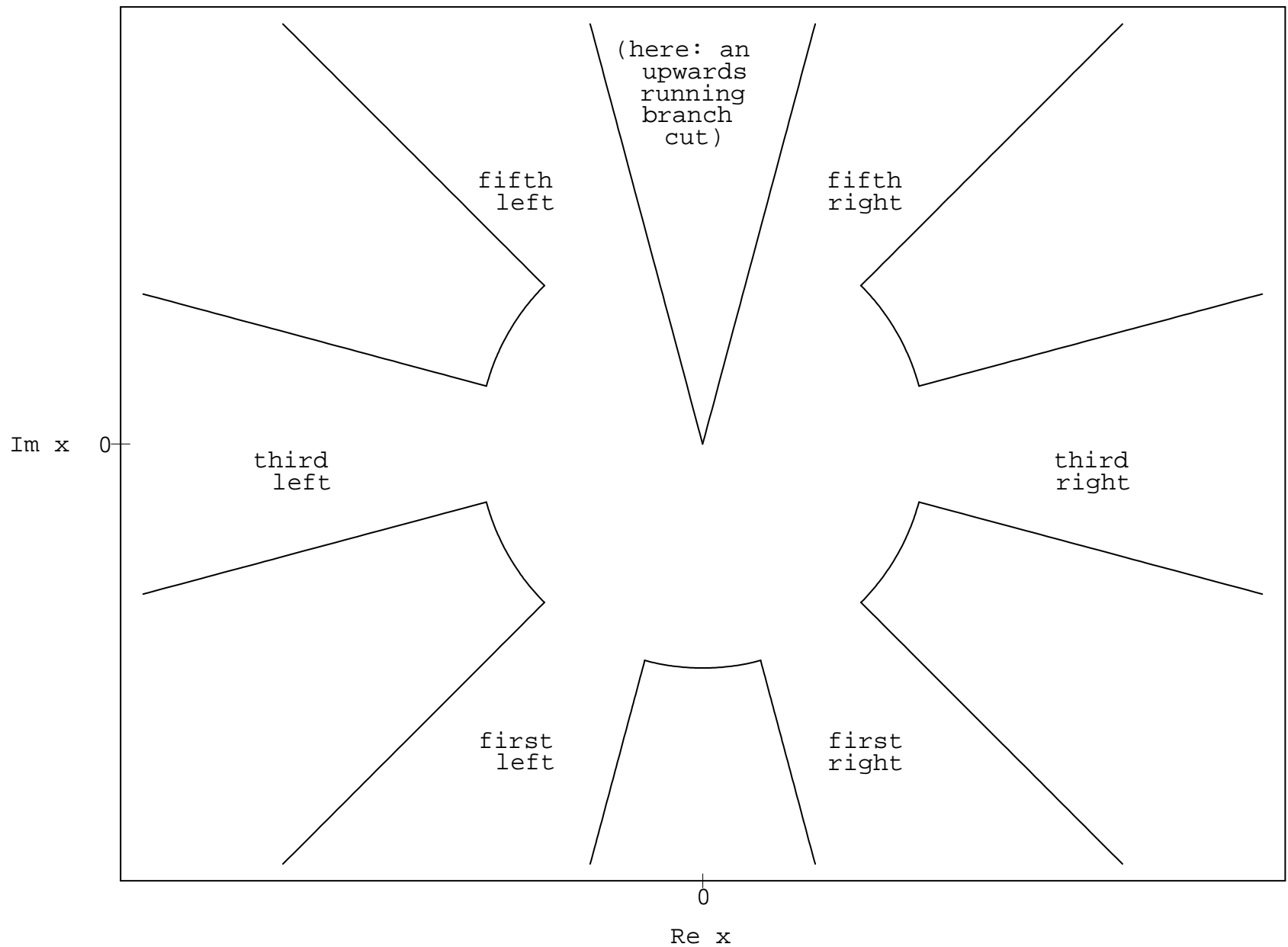
\subsection*{Figure 1. Asymptotic wedges for decadic oscillators.}

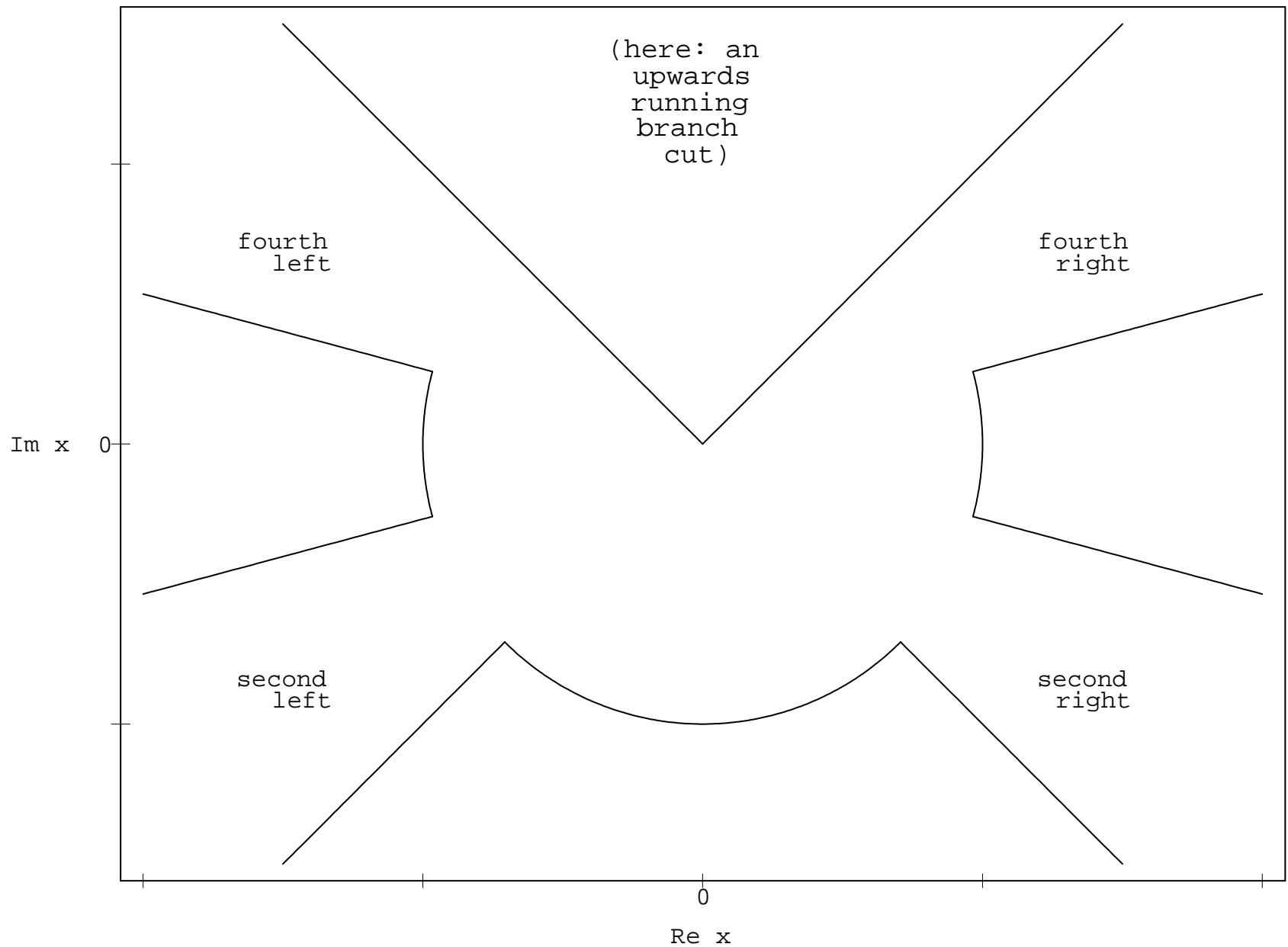
\subsection*{Figure 2. The ``unphysical" alternative to Figure 1.}

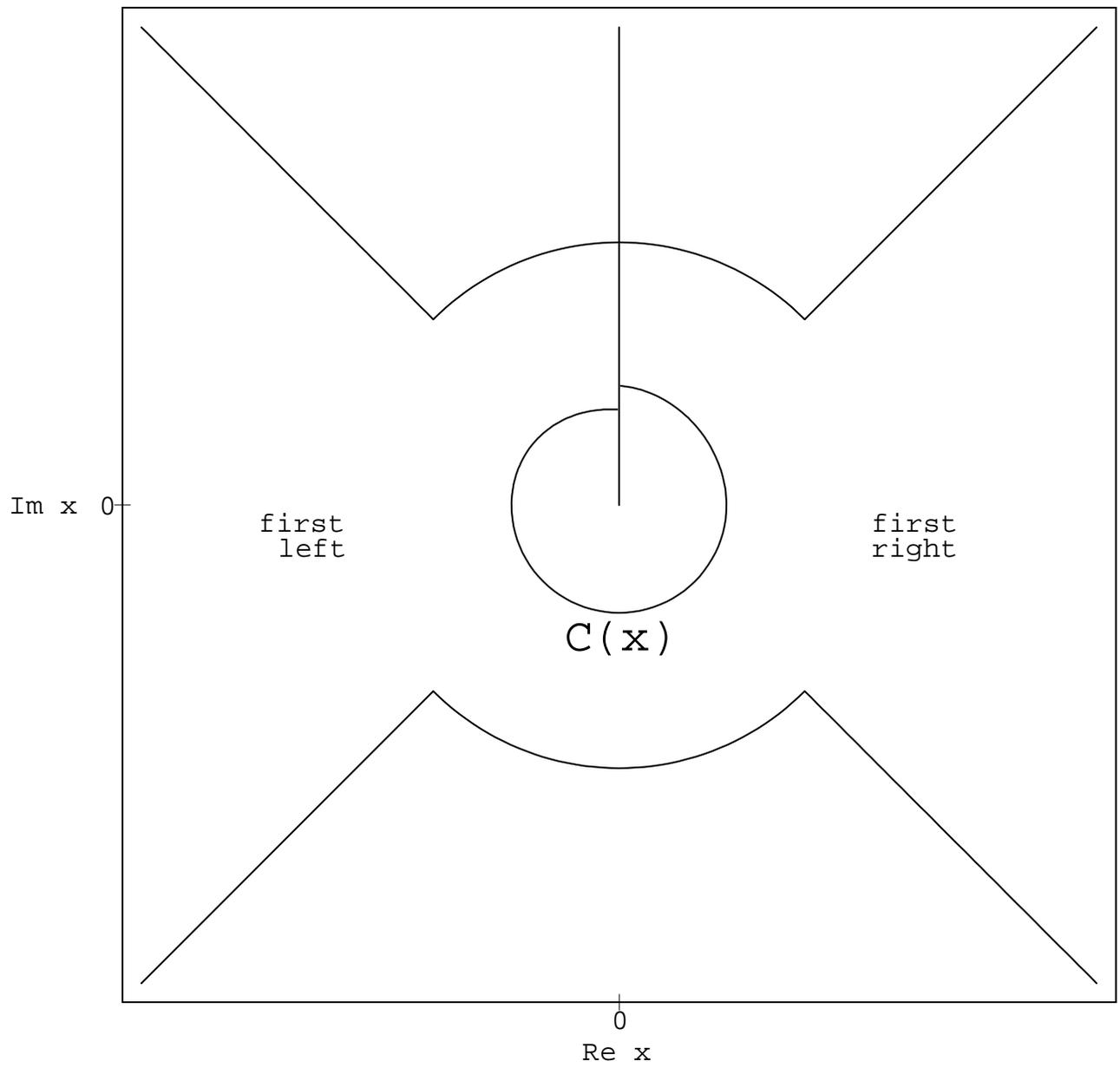
\subsection*{Figure 3. Cut plane with wedges for harmonic oscillator.}

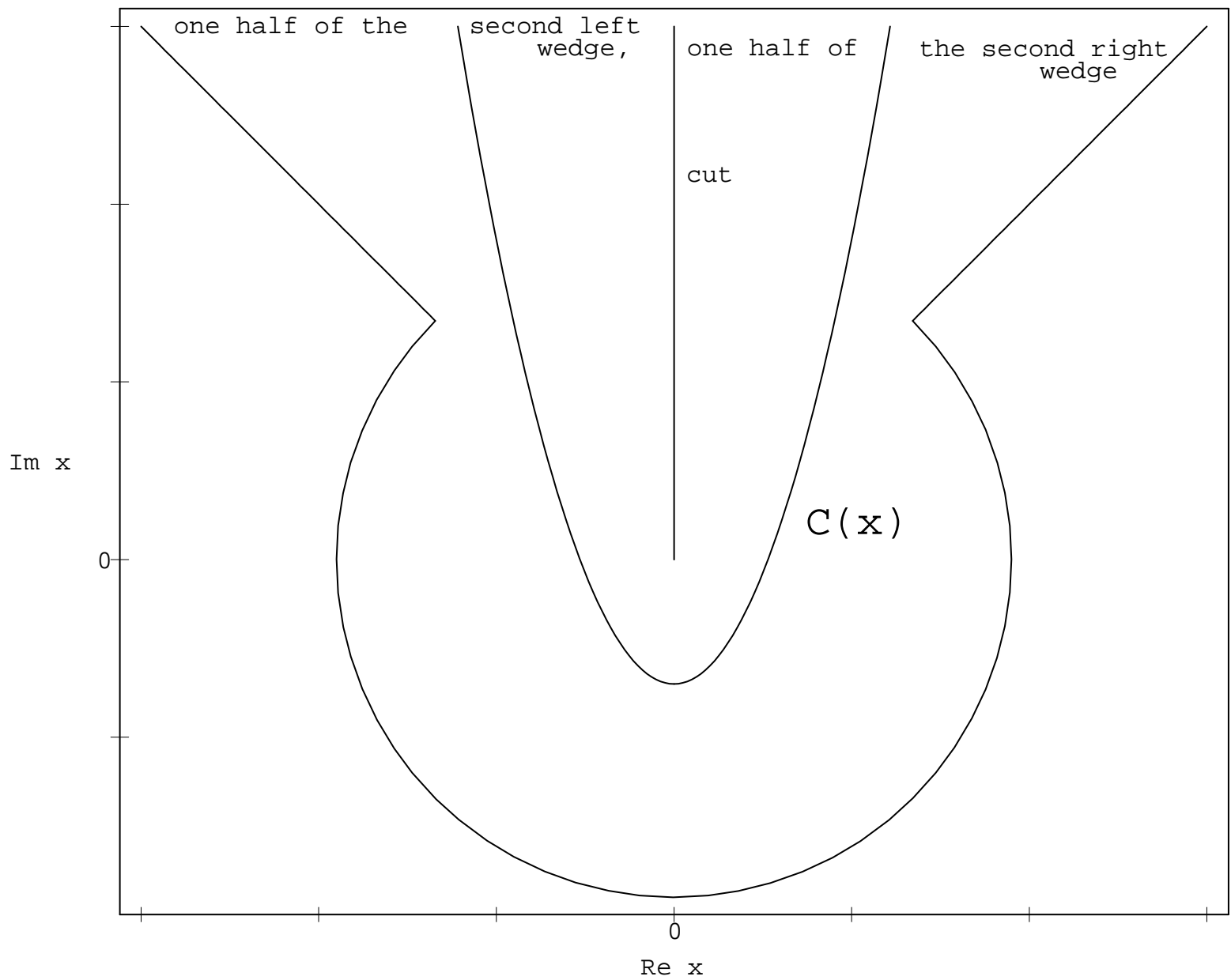
\subsection*{Figure 4. The ``left-second" -- ``right-second" contour $C(s)$}

\subsection*{Figure 5. The ``left" half of the ``third-third"  toboggan}

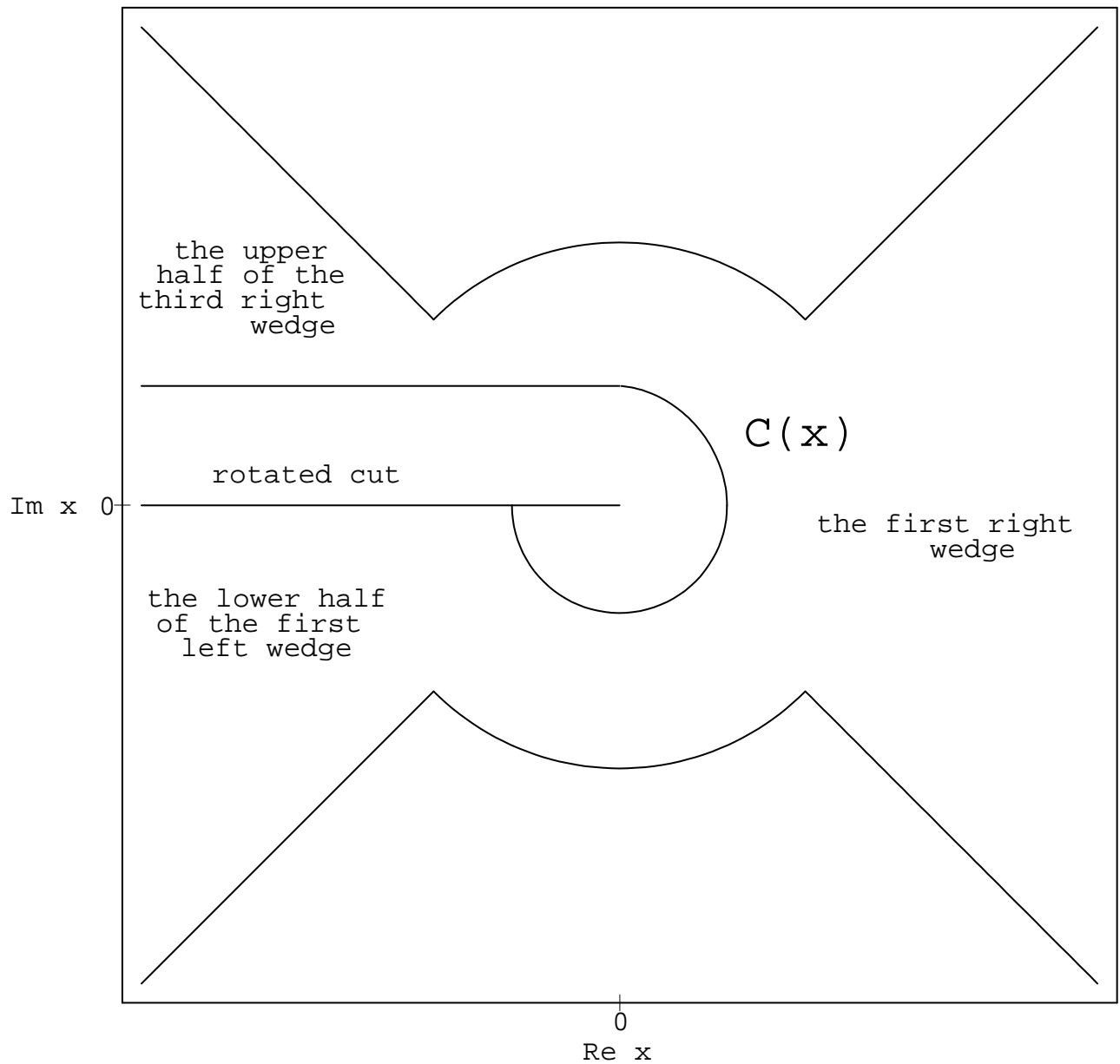
\subsection*{Figure 6. The remaining part of the third-third toboggan.}

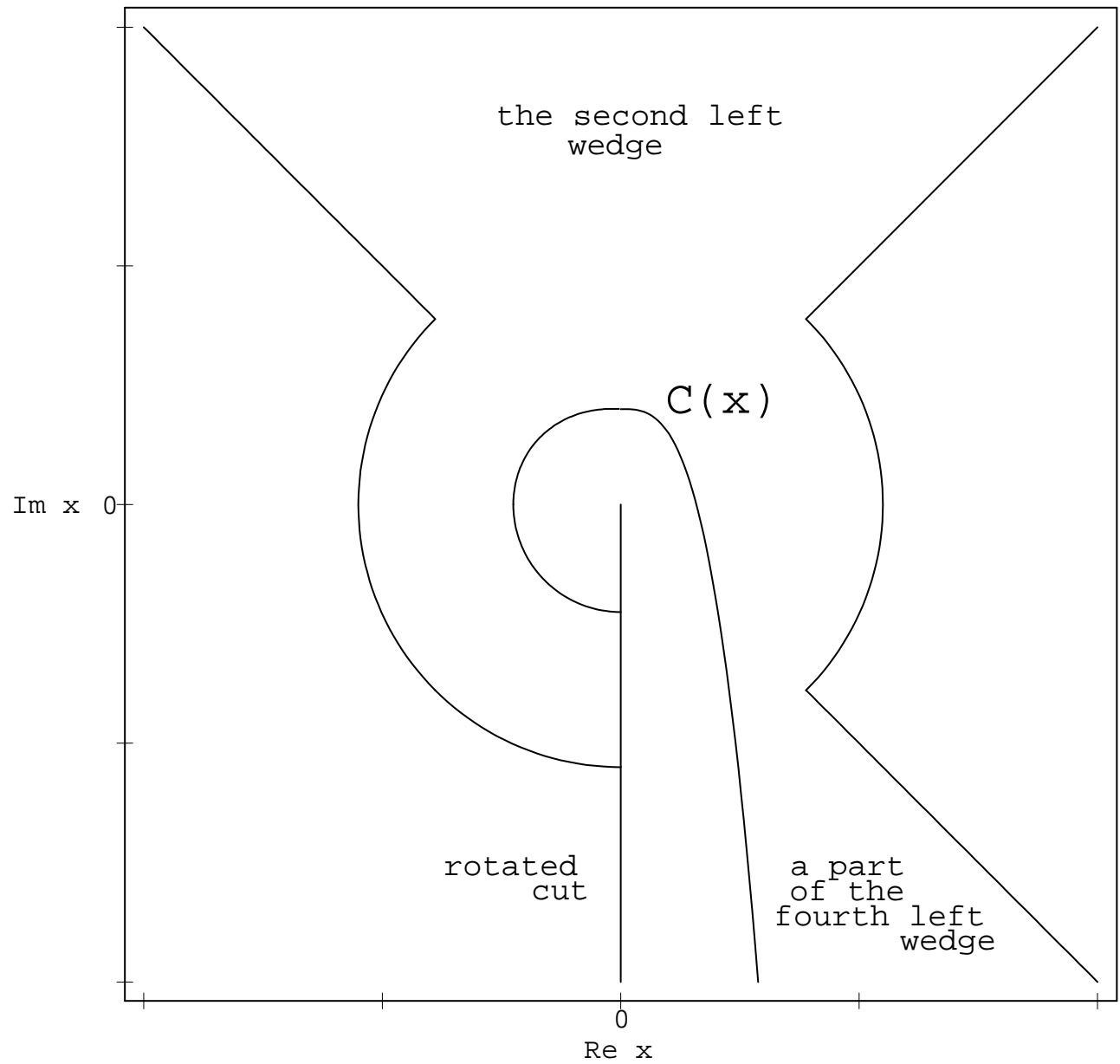
\subsection*{Figure 7. The fourth-fourth toboggan. }

\end{document}